\documentclass[aps,prb,letterpaper,superscriptaddress,citeautoscript,reprint,longbibliography]{revtex4-1}

\usepackage[utf8]{inputenc}
\usepackage{amsmath,amssymb}
\usepackage{graphicx}
\usepackage{hyperref}
\usepackage{color}
\usepackage{bm}

\begin{document}
\title{Excitation of magnon spin photocurrents in antiferromagnetic insulators}

\author{Igor \surname{Proskurin}}
\affiliation{Department of Physics and Astronomy, University of Manitoba, Winnipeg, Manitoba R3T 2N2 Canada}
\email{Igor.Proskurin@umanitoba.ca}
\affiliation{Institute of Natural Sciences and Mathematics, Ural Federal University, Ekaterinburg 620002, Russia}
\affiliation{Chirality Research Centre, Hiroshima University, Higashi-Hiroshima, Hiroshima 739-8526, Japan}

\author{Alexander~S. \surname{Ovchinnikov}}
\affiliation{Institute of Natural Sciences and Mathematics, Ural Federal University, Ekaterinburg 620002, Russia}
\affiliation{Institute for Metal Physics, Ural Division of the Russian Academy of Sciences, Ekaterinburg 620137, Russia}

\author{Jun-ichiro \surname{Kishine}}
\affiliation{Division of Natural and Environmental Sciences, The Open University of Japan, Chiba 261-8586, Japan}

\author{Robert~L. \surname{Stamps}}
\affiliation{Department of Physics and Astronomy, University of Manitoba, Winnipeg, Manitoba R3T 2N2 Canada}

\begin{abstract}
In the circular photogalvanic effect, circularly polarized light can produce a direct electron photocurrent in metals and the direction of the current depends on the polarization. We suggest that an analogous nonlinear effect exists for antiferromagnetic insulators wherein the total spin of light and spin waves is conserved. In consequence, a spin angular momentum is expected to be transfered from photons to magnons so that a circularly polarized electromagnetic field will generate a direct magnon spin current. The direction of the current is determined by the helicity of the light. We show that this resonant effect appears as a second order light-matter interaction. We find also a geometric contribution to the spin photocurrent, which appears for materials with complex lattice structures and Dzyaloshinskii-Moriya interactions.
\end{abstract}

\date{\today}
\maketitle

%% ==================================================================
%% ==================================================================

\section{Introduction}
Antiferromagnetic insulators are promising candidates with which to address the problem of creating and transmitting spin currents--one of the key issues facing spintronics\cite{Jungwirth2016,Gomonay2017,Baltz2018,Smejkal2018}. 
Spin currents in these materials are carried by magnons \cite{Cheng2014, Rezende2016a} and possess nontrivial topological properties \cite{Li2016,Owerre2016,Owerre2018,Jian2018} including chirality and its conservation \cite{Proskurin2017,Daniels2018}. A number of possible applications have been proposed that exploit unique properties of antiferromagnetic magnons  \cite{Cheng2014a,Daniels2015,Sekine2016,Cheng2016,Khymyn2017,Zyuzin2017,Cheng2018}. Magnonic spin currents in antiferromagnets can be created in several ways. There is a spin pumping mechanism from a ferromagnetic layer \cite{Wang2014}, injection from a metallic layer across the interface \cite{Lin2017}, and generation using a temperature gradient via a magnonic spin Seebeck effect \cite{Seki2015, Rezende2016,Wu2016,Holanda2017}. Additionally, a magnonic spin Nernst effect was proposed for quasi-two-dimensional hexagonal antiferromagnets with Dzyaloshinskii-Moriya interactions (DMI) \cite{Cheng2016a,Zyuzin2016} and found later experimentally in MnPS$_{3}$ \cite{Shiomi2017}. Ultrafast optical excitation of coherent magnon dynamics \cite{Satoh2010,Tzschaschel2017,Satoh2017} is another prominent area, which is in the basic concept of antiferromagnetic optospintronics -- a direction targeting optical control of spin states in antiferromagnetic insulators \cite{Nemec2018}.

\begin{figure}
	\centerline{\includegraphics[width=0.415\textwidth]{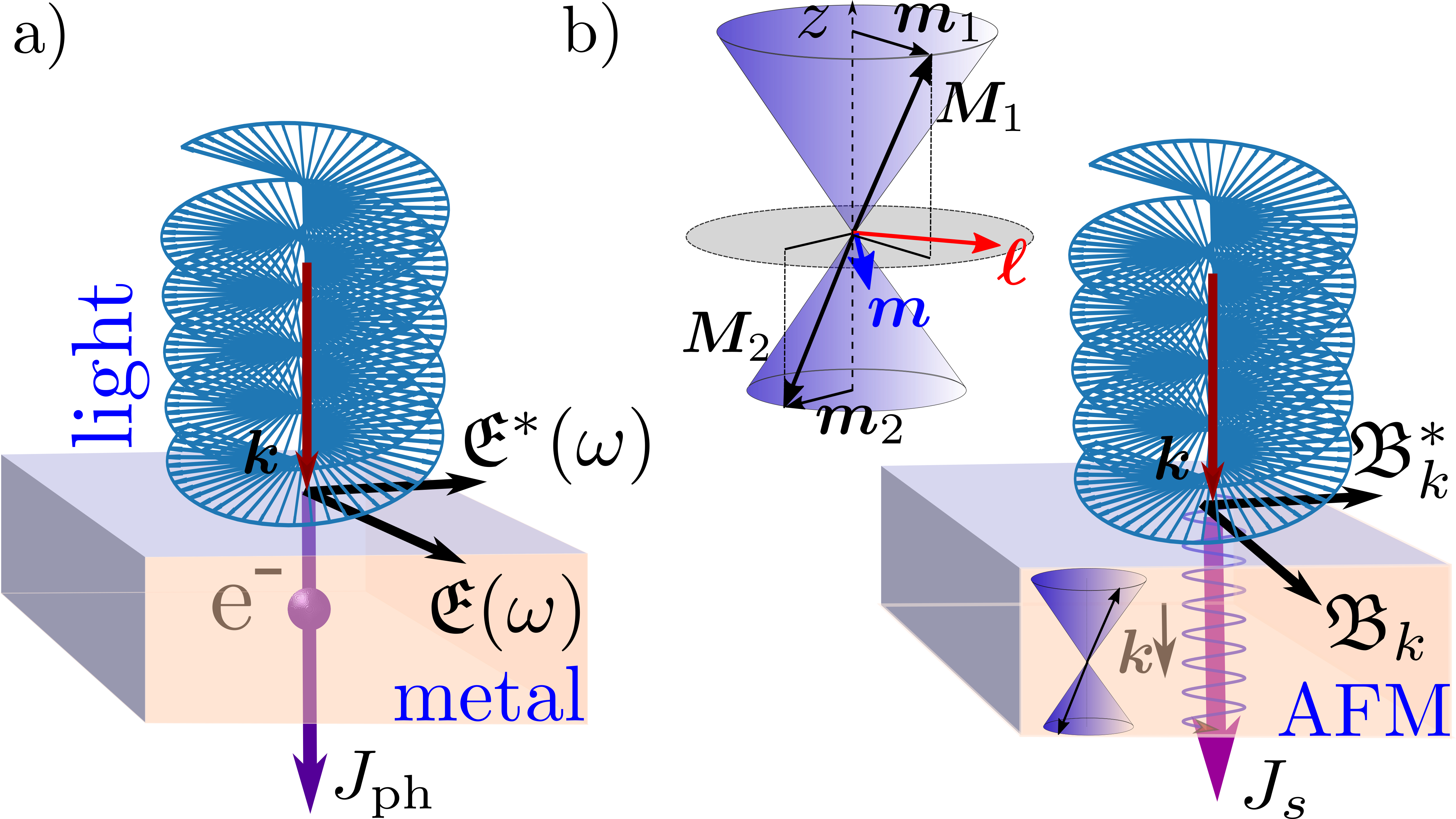}}
	\caption{Schematic picture of the circular photogalvanic effect in metals where $\bm{J}_{\mathrm{ph}} \sim i\bm{\mathfrak{E}}(\omega) \times \bm{\mathfrak{E}}^{*}(\omega)$ is the direct photocurrent generated by circularly polarized light (a), and the proposed optical excitation of the spin current in antiferromagnetic insulators (b). The polarized light with the wave vector $\bm{k}$ is propagating along the magnetic ordering direction  exciting magnon spin current $\bm{J}_{s}$. Precession of the sublattice magnetizations $\bm{M}_{1}$ and $\bm{M}_{2}$ is shown above.}
	\label{fig1}
\end{figure}

Although an additional antiferromagnetic layer in a ferromagnet/normal metal interface can sufficiently enhance efficiency of spin pumping \cite{Wang2014,Khymyn2016}, and dynamic antiferromagnets can themselves serve as sources of spin currents experiencing a spin backflow from adjacent metal slabs  \cite{Johansen2017}, nonthermal generation of magnon spin currents in bulk antiferromagnets is challenging due to vanishing net magnetic moment.
In this paper, we consider a mechanism for generating spin currents in antiferromagetic insulators by optical excitation of spin dynamics. The mechanism involves polarized light and is motivated by the photogalvanic effect. In metals, the circular photogalvanic effect is a nonlinear optical response to the circularly polarized electric field $\bm{\mathfrak{E}}(\omega)$ that generates a direct electron photocurrent $\bm{J}_{\mathrm{ph}} = i\hat{\beta}(\omega) [\bm{\mathfrak{E}}(\omega) \times \bm{\mathfrak{E}}^{*}(\omega)]$, where $\hat{\beta}$ is the material tensor, which is nonzero in metals lacking the inversion symmetry \cite{Belinicher1980}. The photocurrent reverses its direction when the polarization of light is switched (see Fig.~\ref{fig1}~a). 

%% ==================================================================
%% ==================================================================

Similar to metallic systems, a symmetry argument suggests that the magnetic component of a light wave, $\bm{\mathfrak{B}}(\omega)$, propagating in an insulating antiferromagnet is able to create a magnonic spin current proportional to $\bm{\mathfrak{B}}(\omega) \times \bm{\mathfrak{B}}^{*}(\omega)$, as schematically shown in Fig.~\ref{fig1}~(b). The argument goes as follows. The spin current, which is determined by the continuity equation
$\partial s_{i}/\partial t + \bm{\nabla} \cdot \bm{J}_{s}^{(i)} = 0$ for the conserving local spin density component $s_{i}$, is odd under the spatial inversion and even under the time reversal transformations. These symmetry properties are exactly like those of the optical chirality $ C_{\chi}= \int \frac{\varepsilon_{0}}{2} \bm{\mathfrak{E}}\cdot \bm{\nabla} \times \bm{\mathfrak{E}} +\frac{1}{2\mu_{0}} \bm{\mathfrak{B}}\cdot \bm{\nabla} \times \bm{\mathfrak{B}} d\bm{r}$ that recently became a useful quantity in optics and plasmonics  \cite{Tang2010,Tang2011,Hendry2010} for characterizing the asymmetry in light-matter interactions \cite{Bliokh2014,Canaguier2013}. The electromagnetic field with nonzero $ C_{\chi}$ corresponds to unequal numbers of left and right polarized photons \cite{Coles2012}. Excitation of antiferromagnetic magnons by this field will likewise create unequal populations of left and right polarized magnon states. The unequal magnon populations can be interpreted as the transfer of spin angular momentum from the light to the antiferromagnet. The density of optical chirality in the momentum space is given by $ \rho_{\chi}(\bm{k}) = \frac{\varepsilon_{0}}{2} \bm{k} \cdot  [\bm{\mathfrak{E}}_{\bm{k}} \times \bm{\mathfrak{E}}^{*}_{\bm{k}}] + \frac{1}{2\mu_{0}} \bm{k} \cdot  [\bm{\mathfrak{B}}_{\bm{k}} \times \bm{\mathfrak{B}}^{*}_{\bm{k}}]$, and, as a result, we may expect that the chiral field can be a source of spin currents determined by the second term. This effect should be most prominent when the light resonantly couples to the spin system.

In what follows, we support our symmetry arguments by detailed calculations.  Semiclassical analysis is provided in Sec.~\ref{sec2}, and followed by microscopic calculations based on the second-order response theory in Sec.~\ref{sec3}. Section~\ref{sec4} is reserved for the results and discussion, while a short summary is given in Sec.~\ref{sec5}.

\section{Semiclassical theory}
\label{sec2}

Optical generation of spin currents can be illustrated semiclassically. For a cubic antiferromagnet with two sublattices $\bm{M}_{1}$ and $\bm{M}_{2}$, the magnetic energy can be written as
\begin{multline} \label{wen}
W = \int \frac{\alpha}{2} \left [(\bm \nabla \bm M_{1})^2 + (\bm \nabla 
\bm M_{2})^2 \right ] + \alpha' \bm \nabla \bm M_{1} \cdot \bm \nabla \bm M_{2}
\\
+ \delta \bm M_{1} \cdot \bm M_{2} - \frac{\beta}{2} \left( M^{2}_{1z} +  M_{2z}^{2}
\right)d\bm{r},
\end{multline}
where $\delta$, $\alpha$, and $\alpha'$ are the exchange parameters,  and $\beta>0$ is the easy-axis anisotropy, which stabilizes a uniform antiferromagnetic ordering along the $z$-direction \cite{Akhiezer1968}. Semiclassical dynamics of sublattice magnetizations can be described by the Landau-Lifshitz-Gilbert equations of motion,
$\partial \bm{M}_{i}/\partial t = \gamma \bm{M}_{i} \times \bm{H}_{i}^{\mathrm{eff}} + \eta_{G}M_{s}^{-1} \bm{M}_{i}\times\partial \bm{M}_{i}/\partial t$, where $\gamma$ denotes the gyromagnetic ratio, $M_{s}$ is the saturation magnetization, $\eta_{G}$ is the Gilbert damping, and the effective fields are determined as follows, $ \bm{H}_{i}^{\mathrm{eff}} = -\delta W/\delta \bm{M}_{i} $ ($ i = 1,2 $).

For small fluctuations around the ordered state, the equations of motion can be written in a linearized form by introducing fluctuating components of the sublattice magnetizations, $\bm M_{1} = M_{s} \hat{\bm{z}} + \bm{m}_{1}(t,\bm{r})$ and $\bm M_{2} = -M_{s} \hat{\bm{z}} + \bm{m}_{2}(t,\bm{r})$, where $\hat{\bm{z}}$ is the unit vector along the $z$-direction (see Fig.~\ref{fig1}~(b)). By transforming to the momentum space $\bm{m}_{i}(t, \bm{r}) = V^{-1/2}\sum_{\bm{k}} \exp(i\bm{k}\cdot\bm{r}) \bm{m}_{i\bm{k}}(t)$ ($i = 1,2$),  introducing $\bm{m} = \bm{m}_{1\bm{k}} + \bm{m}_{2\bm{k}}$ and $\bm{l}_{\bm{k}} = \bm{m}_{1\bm{k}} - \bm{m}_{2\bm{k}}$, and expanding up to the linear order in $\bm{m}_{\bm{k}}$ and $\bm{l}_{\bm{k}}$, the Landau-Lifshitz-Gilbert equations can be written as
\begin{align}
\frac{\partial \bm{m}_{\bm{k}}}{\partial t}  &= -\varepsilon_{\bm{k}}^{(l)} \hat{\bm{z}} \times \bm{l}_{\bm{k}} + \eta_{G} \hat{\bm{z}} \times \frac{\partial \bm{l}_{\bm{k}}}{\partial t} , \\ \label{eql}
\frac{\partial \bm{l}_{\bm{k}}}{\partial t} &= -\varepsilon_{\bm{k}}^{(m)} \hat{\bm{z}} \times \bm{m}_{\bm{k}}+ \eta_{G} \hat{\bm{z}} \times \frac{\partial \bm{m}_{\bm{k}}}{\partial t} ,
\end{align}
where $\varepsilon_{\bm{k}}^{(l)} = \gamma M_{s}(\beta  + (\alpha - \alpha') k^2)$ and $\varepsilon_{\bm{k}}^{(m)} = \gamma M_{s}(2\delta + \beta + (\alpha + \alpha') k^2)$. The solution of these equations gives a pair of degenerated antiferromagnetic spin waves with opposite polarizations and energy dispersion $\varepsilon_{\bm{k}} = (\varepsilon_{\bm{k}}^{(l)}\varepsilon_{\bm{k}}^{(m)})^{\frac{1}{2}}$.

When the rotational symmetry is preserved for the interaction between the electromagnetic field and the spin system, the expression for the spin current can be found from the equation of motion for the $z$-component of the magnetization density $m^{z} = \frac{1}{2M_{s}}(m_{2}^{2} - m_{1}^{2})$, which can be written as follows
\begin{multline} \label{mzq}
\frac{\partial m^{z}_{\bm{q}}}{\partial t}  = \frac{1}{4M_{s}} \sum_{\bm{k}} \left \lbrace 
\left(\varepsilon^{(l)}_{\bm{k}-\bm{q}} - \varepsilon^{(l)}_{-\bm{k}}\right)  \left[\bm{l}^{*}_{\bm{k}-\bm{q}} \times \bm{l}_{\bm{k}} \right]_{z} 
\right.
\\
\left.
+ \left (\varepsilon^{(m)}_{-\bm{k}+\bm{q}} -\varepsilon^{(m)}_{\bm{k}} \right)\left[\bm{m}^{*}_{\bm{k}-\bm{q}} \times \bm{m}_{\bm{k}} \right]_{z}
\right \rbrace,
\end{multline}
which, taking into account that $\varepsilon^{(\alpha)}_{-\bm{k}} = \varepsilon^{(\alpha)}_{\bm{k}}$ ($\alpha=m,l$), in the long-wavelength limit can be expressed in the form of the continuity equation \footnote{Note that the Gilbert damping produces additional term on the right hand side of Eq.~\eqref{mzq}, $ I_{G} = -\frac{\eta_{G}}{2M_{s}}\sum_{k}\left(\varepsilon^{(m)}_{\bm{k}-\bm{q}}\bm{m}^{*}_{\bm{k}-\bm{q}}\cdot\bm{l}_{\bm{k}} +  \varepsilon^{(l)}_{\bm{k}}\bm{m}^{*}_{\bm{k}-\bm{q}}\cdot\bm{l}_{\bm{k}}\right)$, related to the dissipation of angular momentum through the relaxation of $m^{z}_{\bm{q}}$. In what follows, we imply that the damping is small, so that the description based on the continuity equation remains valid.} $ \partial_{t} m^{z}_{\bm{q}} + i\bm{q} \cdot \bm{J}_{s} = 0$, with
\begin{equation} \label{Jm}
\bm{J}_{s}=\frac{-i}{4M_{s}} \sum_{\bm{k}}  \left[ \frac{\partial \varepsilon_{\bm{k}}^{(m)}}{\partial \bm{k}} (\bm{m}^{*}_{\bm{k}}\times\bm{m}_{\bm{k}})_{z} + \frac{\partial\varepsilon_{\bm{k}}^{(l)}}{\partial \bm{k}} (\bm{l}^{*}_{\bm{k}}\times\bm{l}_{\bm{k}})_{z} \right].
\end{equation}
being the spatially homogeneous magnon spin current.

In order to study light-induced magnetization dynamics, we add the interaction term $ -\bm{h}(t)\cdot(\bm{M}_{1} + \bm{M}_{2}) $ between the magnetic field of the electromagnetic wave $\bm{h}(t)$ and sublattice magnetizations. In the linear equations of motion it corresponds to the additional term $2\gamma M_{s} \hat{\bm{z}}\times\bm{h}_{\bm{k}}$ in Eq.~\eqref{eql}, where $ \bm{h}_{\bm{k}} $ denotes the Fourier component of the magnetic field with the wave vector $\bm{k}$. For $ \bm{h}_{\bm{k}} \sim \exp(i\omega t-ikz) $, we obtain the solutions,
\begin{align}
\bm{m}_{k}(\omega) & = 2\gamma M_{s}\chi_{m}(k,\omega)\bm{h}_{k}(\omega), \\
\bm{l}_{k}(\omega) & = 2i\gamma M_{s}\chi_{l}(k,\omega) [\hat{\bm{z}} \times\bm{h}_{k}(\omega)]_{z},
\end{align}
where the susceptibilities in the absence of dissipation are given by $ \chi_{m}(k,\omega) = \varepsilon_{k}^{(l)}/(\varepsilon_{k}^{2} - \omega^{2}) $ and $ \chi_{l}(k,\omega) = \omega/(\varepsilon^{2}_{k} - \omega^{2}) $. Small damping can be included in $\chi_{i}$ by replacing $ \varepsilon_{\bm{k}}^{(i)}$ with $\varepsilon_{\bm{k}}^{(i)} - i\eta_{G}\omega$ ($i = m,l$). With the help of these expressions, the field-induced spin current along the $z$-direction takes the form
\begin{multline} \label{Js1c}
J_{s}^{(z)} = -i\gamma^{2} M_{s}\sum_{k} \left(
\chi_{m}^{2}(k,\omega)\nabla_{k}\varepsilon^{(m)}_{k} + 
\chi_{l}^{2}(k,\omega)\nabla_{k}\varepsilon^{(l)}_{k}
\right) 
\\ 
\times \left[\bm{h}^{*}_{k}(\omega) \times \bm{h}_{k}(\omega)\right]_{z},
\end{multline}
which takes a maximum in the region $\varepsilon_{\bm{k}} \approx \omega$ where the coupling of light and spin waves is resonantly strong. Since both $ \nabla_{k}\varepsilon^{(m)}_{k} $ and $\nabla_{k}\varepsilon^{(l)}_{k}$ are proportional to $ k $, the asymmetric combination $ k \left[\bm{h}^{*}_{k}(\omega) \times \bm{h}_{k}(\omega)\right]_{z} $ appears at the right-hand side, in agreement with our symmetry arguments. Note that $  J_{s}^{(z)}$ is nonzero only for a wave with finite wave vector, and vanish in the $ k \to 0 $ limit.

Before going to further analysis, we generalize Eq.~\eqref{Js1c} in the next section using a microscopic derivation.

\section{Nonlinear response theory}
\label{sec3}

Deeper insight into optical generation of spin currents can be obtained with a calculation of the spin current using nonlinear response theory. We begin with a Hamiltonian for the antiferromagnet assuming symmetric exchange:
\begin{multline} \label{Hm}
\mathcal{H} = \sum_{\langle ij \rangle} \frac{{J}_{ij}}{2} \left(S_{i}^{(+)}S_{j}^{(-)} +   S_{i}^{(-)}S_{j}^{(+)} \right) \\
 + \sum_{\langle ij \rangle} J_{ij} S_{i}^{z}S_{j}^{z} - K \sum_{i} (S_{i}^{z})^{2},
\end{multline}
which includes symmetric exchange interactions $J_{ij}$ between the first and the second nearest neighboring sites, the single ion anisotropy constant $K \sim \beta a^{-3}$ (in what follows we take the lattice constant $a = 1$). Later we will comment on effects that can arise when a Dzyaloshinski Moriya  interaction appears as an asymmetric contribution to the exchange interaction. 

We now transform the Hamiltonian \eqref{Hm} using a Hostein-Primakoff representation with $S_{iA}^{(+)} = \sqrt{2S}a_{i}$, $S_{iB}^{(+)} = \sqrt{2S}b_{i}^{\dag}$, $S_{iA}^{(-)} = \sqrt{2S}a_{i}^{\dag}$, $S_{iB}^{(-)} = \sqrt{2S}b_{i}$, $S_{iA}^{z} = S - a_{i}^{\dag} a_{i}$, and $S_{iB}^{z} = -S + b_{i}^{\dag} b_{i}$, where $A$ and $B$ are sublattice indexes. Terms to the second order in $a_{i}$ and $b_{i}$ are kept. Using spin wave variables defined as $a_{i} = N^{-1/2}\sum_{\bm{k}} e^{i\bm{k} \cdot \bm{r}} a_{\bm{k}}$ and $b_{i} = N^{-1/2}\sum_{\bm{k}} e^{i\bm{k} \cdot \bm{r}} b_{\bm{k}}$, we rewrite Eq.~\eqref{Hm} as
\begin{equation} \label{Hm1}
\mathcal{H} = \sum_{\bm{k}} \left[A_{\bm{k}}\left(a_{\bm{k}}^{\dag}a_{\bm{k}} + b_{-\bm{k}}^{\dag}b_{-\bm{k}}\right) + B_{\bm{k}}a_{\bm{k}}b_{-\bm{k}} + B^{*}_{\bm{k}}a_{\bm{k}}^{\dag}b_{-\bm{k}}^{\dag}\right].
\end{equation}
Here $A_{\bm{k}} = 2KS + ZJ_{1}S - 2SJ_{2}G_{\bm{k}}$ includes exchange interactions between the first ($J_{1}$) and the second ($J_{2}$) nearest neighbors with the lattice form factor $G_{\bm{k}} = \sum_{\bm{\delta}'} \sin^{2}(\bm{k} \cdot \bm{\delta}')$, and the summation is over next nearest neighbor sites. The parameter $B_{\bm{k}} = |B_{\bm{k}}|\exp(-i\varphi_{\bm{k}})$ contains information about the lattice configuration and intersublattice DMI interactions. Without DMI terms, $B_{\bm{k}} = J_{1}SC_{\bm{k}}$, where the structure factor is given by $C_{\bm{k}} = \sum_{\bm{\delta}} \exp(-i \bm{k} \cdot \bm{\delta})$. The vector $\bm{\delta}$ connects $Z$ nearest neighboring sites.

We consider optical excitation of spin dynamics. Interaction with the electromagnetic field is represented by a Zeeman coupling as $ \mathcal{H}_{I} = - g\mu_{B}\sum_{i} \bm{\mathfrak{B}}(t,\bm{r}_{i}) \cdot \bm{S}_{i}$, where $\mu_{B}$ denotes the Bohr magneton and $g$ is the Land\'{e} factor, as used in cavity electrodynamics for the magnon-photon interaction \cite{Zhang2014a,Cao2015,Yao2015,Sharma2017,Johansen2018}. The interaction term can be rewritten as
\begin{equation} \label{Hi}
\mathcal{H}_{I} = -g\mu_{B}\sqrt{\frac{S}{2}}\sum_{\bm{k}} \left[\mathfrak{B}^{(-)}_{\bm{k}}(t)\left(a_{\bm{k}} + b_{-\bm{k}}^{\dag}\right) + \mbox{h.c.}\right],
\end{equation}
where $\mathfrak{B}^{(\pm)}_{\bm{k}} = \mathfrak{B}^{x}_{\bm{k}} \pm i\mathfrak{B}^{y}_{\bm{k}}$ is the circular Fourier component of the magnetic field defined as $\bm{\mathfrak{B}}(\bm{r}) = N^{-1/2}\sum_{\bm{k}} \exp(-i\bm{k}\cdot\bm{r})\bm{\mathfrak{B}}_{\bm{k}}$. Note that this satisfies the identity $(\mathfrak{B}_{\bm{k}}^{(-)})^{*} = \mathfrak{B}_{-\bm{k}}^{(+)}$. We do not consider coupling between  $S^{z}$ and $\mathfrak{B}^{z}$, since we consider only electromagnetic waves traveling along the $z$-direction. 

To diagonalize Eq.~\eqref{Hm1}, we apply a Bogolyubov transformation using two parameters 
\begin{equation} \label{Bog}
\left(
\begin{array}{c}
a_{\bm{k}} \\
b_{-\bm{k}}^{\dag}
\end{array}
\right)=\left(
\begin{array}{cc}
\cosh \theta_{\bm{k}} e^{i\phi_{\bm{k}}} & -\sinh \theta_{\bm{k}} \\
-\sinh \theta_{\bm{k}} & \cosh \theta_{\bm{k}} e^{-i\phi_{\bm{k}}}
\end{array}
\right)\left(
\begin{array}{c}
\alpha_{\bm{k}} \\
\beta_{-\bm{k}}^{\dag}
\end{array}
\right),
\end{equation}
where $\alpha_{\bm{k}}$ and $\beta_{\bm{k}}$ are operators in the transformed frame. The parameters of the transformation are given by $\tanh 2\theta_{\bm{k}} = |B_{\bm{k}}|/A_{\bm{k}}$, and $\phi_{\bm{k}} = \varphi_{\bm{k}}$. After the transformation, the Hamiltonian in Eq.~\eqref{Hm1} becomes 
\begin{equation} \label{Hmagnon}
\mathcal{H} = \sum_{\bm{k}} \varepsilon_{\bm{k}} \left(\alpha_{\bm{k}}^{\dag}\alpha_{\bm{k}} + \beta_{-\bm{k}}^{\dag}\beta_{-\bm{k}}\right),
\end{equation}
where the energy dispersion relation is given by $\varepsilon_{\bm{k}}=\sqrt{A_{\bm{k}}^{2} - |B_{\bm{k}}|^{2}}$.

We now define the magnon spin current. Because the $z$-component of the total spin is a conserved quantity, the local magnon density $n(\bm{r}_{i}) = \sum_{\bm{\delta}}b^{\dag}_{i+\delta}b_{i+\delta} - a^{\dag}_{i}a_{i}$ should satisfy a continuity equation. Similar to the semiclassical analysis of the previous section, the equation of motion for the  $\bm{q}$th Fourier component of the magnon density can be expressed in the long-wave length limit as $\partial n_{\bm{q}}/\partial t + i \bm{q} \cdot \bm{J}_{s} = 0$ (see Appendix~\ref{ap1}), where 
\begin{multline} \label{JJs}
\bm{J}_{s} = \sum_{\bm{k}}\left[
\frac{\partial A_{\bm{k}}}{\partial \bm{k}} \left(a^{\dag}_{\bm{k}} a_{\bm{k}} + 
b^{\dag}_{-k}b_{-k}\right) \right. \\
\left.
+\frac{\partial B_{\bm{k}}}{\partial \bm{k}} a_{\bm{k}} b_{-\bm{k}} +
\frac{\partial B^{*}_{\bm{k}}}{\partial \bm{k}} a^{\dag}_{\bm{k}} b^{\dag}_{-\bm{k}}
\right]
\end{multline}
is the magnon spin current. Note that for cubic antiferromagnets in the continuous approximation, this expression is in agreement with  Eq.~\eqref{Jm}.

Transforming according to Eq.~\eqref{Bog}, we express Eq.~\eqref{JJs} as
\begin{equation} \label{JJs1}
\bm{J}_{s} = \sum_{\bm{k}} \left(\alpha_{\bm{k}}^{\dag},\beta_{-\bm{k}}\right) \left(
\begin{array}{cc}
\bm{\nabla}_{\bm{k}} \varepsilon_{\bm{k}} & \bm{K}_{\bm{k}}^{*} \\
\bm{K}_{\bm{k}} & \bm{\nabla}_{\bm{k}} \varepsilon_{\bm{k}} 
\end{array} \right)
\left(
\begin{array}{c}
\alpha_{\bm{k}} \\
\beta_{-\bm{k}}^{\dag}
\end{array}
\right),
\end{equation}
where the off-diagonal matrix elements are given by
\begin{equation} \label{Kk}
\bm{K}_{\bm{k}} = e^{i\varphi_{\bm{k}}} \left(\frac{A_{\bm{k}}\bm{\nabla}_{\bm{k}}|B_{\bm{k}}| - |B_{\bm{k}}|\bm{\nabla}_{\bm{k}}A_{\bm{k}}}{\sqrt{A_{\bm{k}}^{2} - |B_{\bm{k}}|^{2}}} - i|B_{\bm{k}}|\bm{\nabla}_{\bm{k}}\varphi_{\bm{k}}\right).
\end{equation}
These expressions show two contributions to the spin current. The first is given by diagonal terms and is proportional to the magnon group velocity. The second contribution is from the off-diagonal elements and describes intersublattice dynamics. This contribution contains information about magnon phase.

Considering the interaction with electromagnetic field in Eq.~\eqref{Hi} as a perturbation, optically excited spin current can be calculated using a second-order Kubo response formula \cite{Tiablikov2013}
\begin{multline} \label{Js}
\langle \bm{J}_{s}(t) \rangle = -\sum_{\omega_{1}\omega_{2}} \int_{-\infty}^{t} dt_{1} \int_{-\infty}^{t_{1}} dt_{2}e^{\epsilon(t_{1}+t_{2}-t)} e^{i\omega_{1}t_{1} + i\omega_{2}t_{2}} \\
\times\langle [[\tilde{\bm{J}}_{s}(t),\tilde{\mathcal{H}}_{I}^{(\omega_{1})}(t_{1})],\tilde{\mathcal{H}}_{I}^{(\omega_{2})}(t_{2})]\rangle.
\end{multline}
Here, the average is taken with respect to the density matrix of noninteracting system $\rho = \exp(-\mathcal{H}/k_{B}T)$, where $T$ is the temperature and $k_{B}$ is Boltzmann's constant. All the operators are taken in the Heisenberg picture, $\tilde{\bm{J}}_{s}(t) = \exp(i\mathcal{H}) \bm{J}_{s} \exp(-i\mathcal{H})$ and $\tilde{\mathcal{H}}_{I}^{(\omega)}(t) = \exp(i\mathcal{H}) \mathcal{H}_{I}^{(\omega)} \exp(-i\mathcal{H})$, where $\mathcal{H}_{I}^{(\omega)}$ is the interaction term in the transformed frame. This term is given by
\begin{equation}
\mathcal{H}_{I}^{(\omega)} = \frac{1}{2}\sum_{\bm{k}}\left[h_{\bm{k}}^{(-)}(\omega)\left(M_{\bm{k}}^{*}\alpha_{\bm{k}}+ M_{\bm{k}}\beta^{\dag}_{-\bm{k}}\right) + \mbox{h.c.}\right],
\end{equation}
with $M_{\bm{k}} = \cosh\theta_{\bm{k}}e^{-i\varphi_{\bm{k}}} - \sinh\theta_{\bm{k}}$. Here we introduce a short hand notation $h^{(\pm)}_{\bm{k}}(\omega) = -g\mu_{B}\sqrt{2S} \mathfrak{B}^{(\pm)}_{\bm{k}}(\omega)$.

The spin photocurrent can be obtained from Eq.~\eqref{Js} as follows (see Appendix~\ref{ap2})
\begin{widetext}
\begin{multline} \label{JJss3}
\langle\bm{J}_{s}\rangle = \frac{1}{4}\sum_{\bm{k}} 
\left\lbrace 
|M_{\bm{k}}|^{2} \bm{\nabla}_{\bm{k}} \varepsilon_{\bm{k}} 
\left[ 
\frac{1}{(\varepsilon_{\bm{k}} + \omega_{\bm{k}})^{2} + \epsilon^{2}} +
\frac{1}{(\varepsilon_{\bm{k}} - \omega_{\bm{k}})^{2} + \epsilon^{2}}
\right]
\right.
\\
\left.
+ \frac{\bm{K}_{\bm{k}}M_{\bm{k}}^{2}}{(\varepsilon_{\bm{k}}-\omega_{\bm{k}}-i\epsilon)(\varepsilon_{\bm{k}}+\omega_{\bm{k}}-i\epsilon)} + 
\frac{\bm{K}^{*}_{\bm{k}}M_{\bm{k}}^{*2}}{(\varepsilon_{\bm{k}}-\omega_{\bm{k}}+i\epsilon)(\varepsilon_{\bm{k}}+\omega_{\bm{k}}+i\epsilon)}
\right\rbrace
h^{(-)}_{\bm{k}}(\omega_{\bm{k}})h^{(+)}_{-\bm{k}}(-\omega_{\bm{k}}),
\end{multline}	
\end{widetext}
where $\epsilon$ can be phenomenologically attributed to small magnon damping.

Taking the limit $\epsilon \to 0^{+}$, we find that, similarly to Eq.~\eqref{JJs}, the current consists of two terms, $ \langle \bm{J}_{s} \rangle = \langle \bm{J}_{s}^{(1)} \rangle + \langle \bm{J}_{s}^{(2)} \rangle$. The first term is proportional to the magnon group velocity, $ \bm{v}_{\bm{k}} = \bm{\nabla}_{\bm{k}}\varepsilon_{\bm{k}}$, describing wave packet propagation:
\begin{equation} \label{Js1}
\langle \bm{J}_{s}^{(1)} \rangle = \frac{1}{4} \sum_{\bm{k}} \frac{|M_{\bm{k}}|^{2} (\varepsilon_{\bm{k}}^{2} + \omega^{2})\bm{v}_{\bm{k}}}{(\varepsilon_{\bm{k}}^{2} - \omega^{2})^{2}}
h_{\bm{k}}^{(-)}(\omega)h_{-\bm{k}}^{(+)}(-\omega),
\end{equation}
The second term is related to the fast oscillating intersublattice dynamics in Eq.~\eqref{JJs}:
\begin{equation} \label{Js2}
\langle \bm{J}_{s}^{(2)} \rangle = \frac{1}{4}\sum_{\bm{k}} \frac{\Re(\bm{K}_{\bm{k}} M_{\bm{k}}^{2})}{\varepsilon_{\bm{k}}^{2} - \omega^{2}} h_{\bm{k}}^{(-)}(\omega)h_{-\bm{k}}^{(+)}(-\omega),
\end{equation}
where the explicit expressions for the coefficients read
\begin{align} 
|M_{\bm{k}}|^{2} &= \frac{A_{\bm{k}} - |B_{\bm{k}}|\cos \phi_{\bm{k}}}{\sqrt{A_{\bm{k}}^{2} - |B_{\bm{k}}|^{2}}}, \\\label{M2}
M_{\bm{k}}^{2} &= e^{-i\varphi_{\bm{k}}} \left(\frac{A_{\bm{k}}\cos \varphi_{\bm{k}} - |B_{\bm{k}}|}{\sqrt{A_{\bm{k}}^{2} - |B_{\bm{k}}|^{2}}}-i\sin\varphi_{\bm{k}}\right).
\end{align}
In the most common situation when $\varepsilon_{\bm{k}} = \varepsilon_{-\bm{k}}$,
both $\bm{v}_{\bm{k}}$ and $\bm{K}_{\bm{k}}$ are odd functions of $\bm{k}$, and the only nonzero contribution in Eqs.~\eqref{Js1} and \eqref{Js2} comes from the asymmetric part of the field intensity, $ h_{\bm{k}}^{(-)}(\omega)h_{-\bm{k}}^{(+)}(-\omega) \to i[\bm{h}_{\bm{k}}^{*}(\omega) \times \bm{h}_{\bm{k}}(\omega)]_{z}$. As mentioned above, this quantity is proportional to the difference between the number of left and right polarized photons.

For the case of magnon dynamics with $\varphi_{\bm{k}} = 0$, we can combine Eqs.~(\ref{Js1})--(\ref{M2}) in the following form
\begin{equation}\label{Js3}
\langle \bm{J}_{s} \rangle = \frac{i}{2} \sum_{\bm{k}} \frac{q_{\bm{k}}^{2} \bm{\nabla}_{\bm{k}} p_{\bm{k}} + \omega^{2} \bm{\nabla}_{\bm{k}} q_{\bm{k}}}{(\varepsilon_{\bm{k}}^{2} - \omega^{2})^{2}} [\bm{h}_{\bm{k}}^{*}(\omega) \times \bm{h}_{\bm{k}}(\omega)]_{z},
\end{equation}
where $p_{\bm{k}} = A_{\bm{k}} + |B_{k}|$ and $q_{\bm{k}} = A_{\bm{k}} - |B_{k}|$. This is in agreement with the semiclassical expression in Eq.~\eqref{Js1c} if we identify $p_{\bm{k}}$ with $ \varepsilon^{(m)}_{\bm{k}}$ and $q_{\bm{k}}$ with $ \varepsilon^{(l)}_{\bm{k}}$. 

While intersublattice dynamics is not specific to $\bm{J}_{s}^{(1)}$ and, in general, this terms is related to the ballistic transport of magnons carrying spin angular momentum excited by polarized light, $\bm{J}_{s}^{(2)}$ contains contributions coming essentially from the antiferromagnetic interactions, such as geometric terms related to the magnon phase. In the next section, we discuss the case with $ \partial \varphi_{\bm{k}}/\partial \bm{k} \ne 0 $ and estimate phase contributions to the spin current.

\section{Results and discussion}
\label{sec4}
We analyze the expression for spin current in Eq.~\eqref{Js3}. The gradient terms are proportional to the exchange interactions, $ \bm{\nabla}_{\bm{k}} B_{\bm{k}} \sim J_{1}\bm{k}$ and $ \bm{\nabla}_{\bm{k}} A_{\bm{k}} \sim J_{2}\bm{k}$. Since $J_{1} \gg J_{2}$, the dominant contribution is from the antiferromagnetic exchange interaction, and the spin current is proportional to $J_{1}$. For the case of cubic symmetry, $A_{\bm{k}} \approx 2KS + 6J_{1}S$ and $B_{\bm{k}} \approx 6J_{1}S -J_{1}Sk^{2}$, and we estimate $ \langle \bm{J}_{s}^{(z)} \rangle = 2J_{1}S\sum_{k_{z}}(\chi_{m}^{2} - \chi_{l}^2) k_{z} [\bm{h}_{\bm{k}}^{*}(\omega) \times \bm{h}_{\bm{k}}(\omega)]_{z}$. We note a compensation point when $\omega \approx q_{k}$, estimated as $\omega \approx 2K/\hbar \approx 10^{10}$~s$^{-1}$ for $K \approx 0.1$~K, where the spin current changes the sign. This value, however, may vary considerably for different materials, since the anisotropy is a material-dependent parameter. At high frequencies, away from the resonance $ \omega \gtrsim \varepsilon_{k} $, the magnitude of spin current generated by circularly polarized light can be estimated as $2g^{2}\mu_{B}^{2}J_{1}|\mathfrak{B}|^{2}/(\hbar^{2}\omega c a)$, where $c$ denotes the speed of light. For typical parameters and frequencies in the terahertz range this quantity remains  $\lesssim1$~A/m$^{2}$ for practical field intensities.

The most interesting region for experiment occurs near the antiferromagnetic resonance, $\omega \approx \varepsilon_{\bm{k}}$. In this frequency region, dissipation plays a crucial role, and can be included in our formalism phenomenologically by the replacement $\varepsilon_{\bm{k}} - \omega \pm i\epsilon \approx \pm i \Gamma$ and $\varepsilon_{\bm{k}} + \omega \pm i\epsilon \approx 2\omega_{\mathrm{rs}} \pm i \Gamma$ in Eq.~\eqref{JJss3}, where $\omega_{\mathrm{rs}}$ is the resonant frequency and $\Gamma$ denotes the spin-wave damping. Ballistic transport will occur in materials with small damping and large resonant frequencies, $\omega_{\mathrm{rs}} \gg \Gamma$, where the dominant contribution comes from the term proportional to group velocity. Near the resonance, the spin current is given by
\begin{equation}\label{Js4}
\langle \bm{J}_{s} \rangle \approx   \frac{i q_{\bm{k}}}{4\hbar\omega_{\mathrm{rs}}}  \frac{\bm{v}_{\bm{k}}}{\Gamma^{2}} \left[\bm{h}_{\bm{k}}^{*} \times \bm{h}_{\bm{k}}\right]_{z}.
\end{equation}

We note that experiments have demonstrated \cite{Lin2016} that a thin NiO or CoO layer ($\sim 1$~nm) provides a considerable enhancement of spin current transmission in a multilayer system \cite{Takei2015,Khymyn2016}. This motivates making an estimate of the magnitude of currents that may be expected for the present mechanism for NiO. The magnitude of the resonant spin current can be estimated as $\langle J_{s} \rangle \approx \chi g^{2} \mu_{B}^{2} J_{1}S^{2} c_{s} I_{B} / (2a c^{2} \hbar^{2}\eta_{G}^{2}\omega_{\mathrm{rs}})$, where $\chi = \pm 1$ is the polarization helicity,  $c_{s}$ is the velocity of spin waves, $ I_{B} = |\mathfrak{B}(\omega)|^{2} $ is the intensity of magnetic field, and we take $\Gamma = \eta_{G}\hbar\omega_{\mathrm{rs}}$. For a typical antiferromagnetic insulator such as NiO, we assume $c_{s} = 3 \times 10^{4}$~m/s, $ J_{1} = 200 $~K, $\omega_{\mathrm{rs}} = 30$~THz, $\eta_{G} = 10^{-4}$, $a=0.5$~nm, which gives $\langle J_{s} \rangle \approx 1.5\times10^{4}$~A/m$^{2}$ (in electric units $e/\hbar$) for magnetic field $\mathfrak{B} \approx 10$~mT. Sucha  magnetic field corresponds to the electric field strength of the laser beam $\approx30$~kV/cm, which is below the maximum field strength achieved in THz laser pulses \cite{Hoffmann2009,Hirori2011}. For a focused spot size about 100~$\mu$m the total spin current trough the spot area will be $ \approx0.1$~mA, which is the same order as the current estimated for the magnon Nernst effect \cite{Cheng2016a,Zyuzin2016}.

Lastly, we discuss the effects of DMI and magnon geometrical phase. Generally, with DMI we can excite the spin current even with linearly polarized light. For illustration, we consider a two dimensional antiferromagnet where electromagnetic wave polarized in the $x$-direction is traveling along the $y$-direction with a wave number $k$. There are many possibilities of DMI configurations in a two-dimensional (2D) system, with some being summarized in Ref.~\onlinecite{Kawano2018}. We describe these configurations by modifying the fist term in the Hamiltonian such that a phase term and effective interaction appear, $J_{ij} \to \tilde{J}_{ij}\exp(i\varphi_{ij}) $. We choose $D_{ij}$ to point along the $\hat{\bm{z}}$ direction, and include phase factors with $\tan \varphi_{ij} = D_{ij}/J_{ij}$, with an effective exchange parameter $\tilde{J}_{ij} = (J_{ij}^{2} + D_{ij}^{2})^{1/2}$.

 Let us first consider the case of uniform DMI  on a square lattice, $\sum_{\langle ij \rangle} D_{ij} (\bm{S}_{i} \times \bm{S}_{j})_{z}$, where $D_{ij} = D_{1}$ for nearest neighboring $i$ and $j$ along the $x$-direction, see Fig 2 (a). Such a configuration does not give a complex phase in Eq.~\eqref{Hm1}, but it does shift the origin of $\varepsilon_{\bm{k}}$  by $Q \sim D_{1}/J_{1}$, which leads to a finite group velocity $v_{x} \sim D_{1}$ at $k_{x} = 0$. Similarly to Eq.~\eqref{Js4}, linearly polarized light in the resonant region induces a spin current in the $x$-direction proportional to the intensity of magnetic field $\langle J_{s,x}^{(\varphi)} \rangle \approx  q_{k}v_{x}/(4\hbar\omega_{\mathrm{rs}}\Gamma^{2}) |h^{(x)}_{k}(\omega_{\mathrm{rs}})|^{2}$. The magnitude of this effect will be typically $ D_{1}/J_{1} \approx 10^{-3} $ times smaller than estimated above for circularly polarized field. We note that a similar result is reported in Ref.~\onlinecite{Okuma2018}.

%% ==================================================================
\begin{figure}
	\centerline{\includegraphics[width=0.375\textwidth]{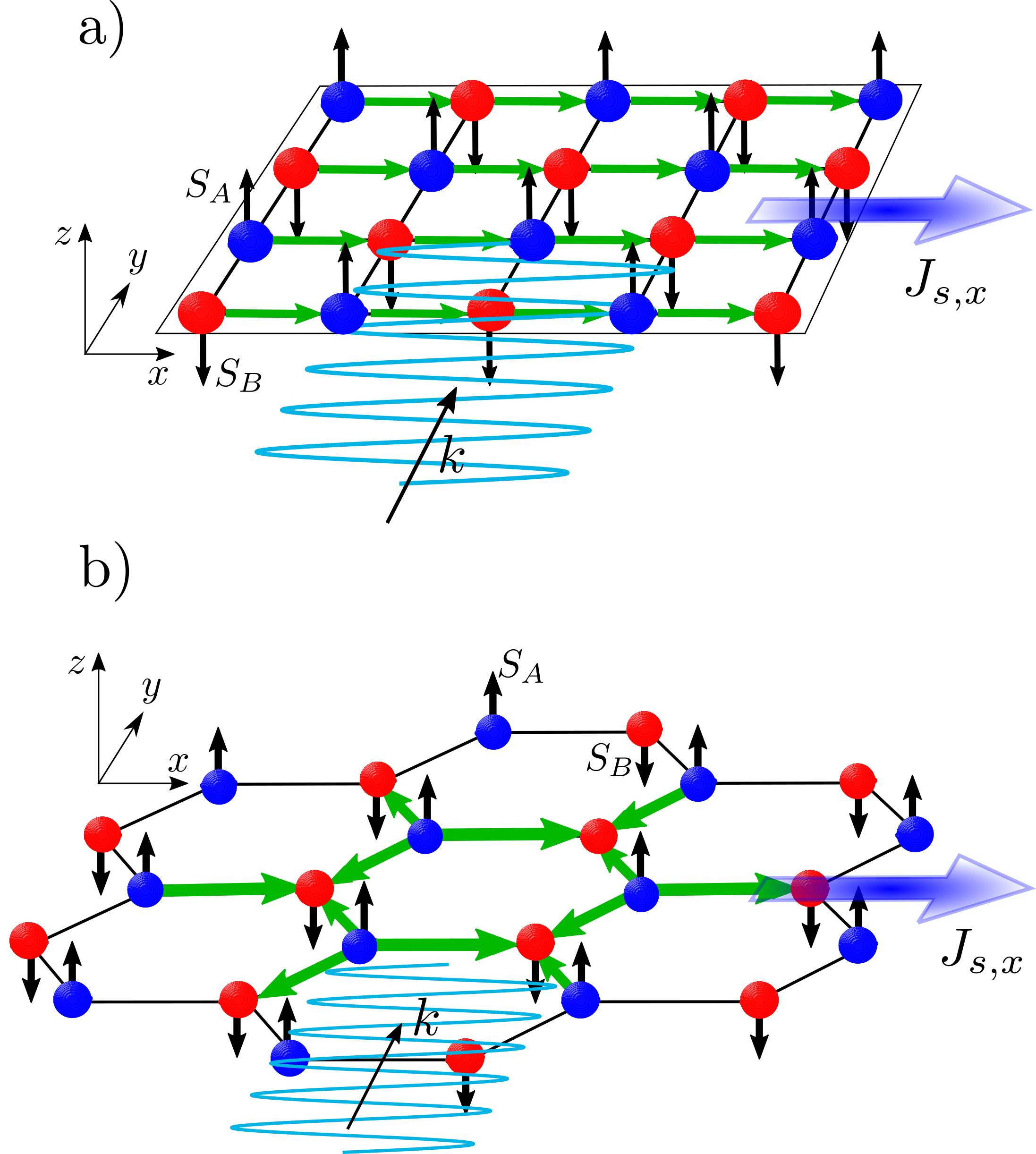}}
	\caption{(a) Schematic picture of 2D antiferromagnet on a square lattice with uniform DMI, $D_{ij} (\bm{S}_{i} \times \bm{S}_{j})_{z}$.  The sign of $D_{ij}$  is positive for $i \to j$ in the direction of green arrows. Spin dynamics is excited by a linearly polarized wave traveling along the $y$-direction, which generates $J_{s,x}$; (b) The same setup for 2D antiferromagnet on a honeycomb lattice with staggered DMI. The sign of $D_{ij}$ is positive for $ i \to j $ pointing from $ A $ to $ B $ sites (marked by the arrows).}
	\label{fig2}
\end{figure}
%% ==================================================================

Equation~\eqref{Js2} shows that there is also a geometrical contribution to the spin current from the phase gradient term in Eq.~\eqref{Kk}, which is given by
\begin{equation}\label{Jsf}
\langle \bm{J}_{s}^{(\varphi)} \rangle =  \frac{1}{2}\sum_{\bm{k}} \frac{|B_{\bm{k}}|\sin \varphi_{\bm{k}} \bm{\nabla}_{\bm{k}} \varphi_{\bm{k}}}{\omega^{2} -\varepsilon_{\bm{k}}^{2}} h_{\bm{k}}^{(-)}(\omega)h_{-\bm{k}}^{(+)}(-\omega).
\end{equation}
This phase $\varphi_{\bm{k}}$ is an offset between the dynamics of the $A$ and $B$ magnetic sublattices, $a_{\bm{k}}(t) \sim \exp(i\varepsilon_{\bm{k}}t)$ and $b^{\dag}_{-\bm{k}}(t) \sim \exp(i\varepsilon_{\bm{k}}t-i\varphi_{\bm{k}})$, owing to the effect of DMI \cite{Kawano2018}. Alternatively, the magnon phase can be generated by the electric field, through the the Aharonov-Casher effect, which was proposed in Refs.~\onlinecite{Owerre2017,Nakata2017} to realize topological magnonic states. This mechanism opens a possible way to manipulate the spin current in Eq.~\eqref{Jsf} by the electric field.

To demonstrate phase effects, let us take the example of a two-dimensional antiferromagnet on a honeycomb lattice. Even without DMI, this model is characterized by finite phase \cite{Cheng2016a,Zyuzin2016}, which satisfies the symmetry condition $\varphi_{\bm{k}} = -\varphi_{-\bm{k}}$. We break this symmetry by a constant phase originating from the nearest neighboring staggered DMI, $D_{ij} = D$ for $ ij=AB $ and $-D$ for $ ij = BA $ (see Fig.~2 (b)). In this case, we have $B_{\bm{k}} = \tilde{J}SC_{\bm{k}}\exp(i\varphi_{0})$, where $\tilde{J} = (J_{1}^{2} + D^{2})^{1/2}$, $\tan \varphi_{0} = D/J_{1}$, and $C_{\bm{k}} = 2\cos(k_{x}/2)\cos(\sqrt{3}k_{y}/2) -1 + 2i\sin(k_{x}/2)[\cos(k_{x}/2) - \cos(\sqrt{3}k_{y}/2)]$ is the structure factor for the honeycomb lattice. In the long wavelength limit we can approximate $B_{\bm{k}} = \tilde{J}S \exp(i\varphi_{0} + i\varphi_{\bm{k}})$, where $\varphi_{\bm{k}} \approx k_{x}(3k_{y}^{2} - k_{x}^{2})/8$. Taking the phase gradient $\nabla_{k_{x}} \varphi_{\bm{k}}$ in the $k_{x} \to 0$ limit, we can generate a spin current perpendicular to direction of wave propagation with a direction controlled by the sign of $\phi_{0}$, i.~e. $\langle J_{s,x}^{(\varphi)} \rangle = 3g^{2}\mu_{B}^{2}\tilde{J}S/(8\hbar^{2} c^{2}) \sin\phi_{0} I_{B} \omega^{2}/(\omega^{2} - \varepsilon_{\bm{k}}^{2})$. Away from the resonance, this expression will be independent of $\omega$.

The phase contributions may be relevant in such quasi-two-dimensional honeycomb materials as MnPS$_{3}$ \cite{Wildes1998} and BiMn$_{4}$O$_{12}$(NO$_{3}$) \cite{Okubo2012}, although these materials have different configurations of DMI terms than assumed here.

\section{Summary}
\label{sec5}

In summary, we have described a resonant induced spin photocurrent for insulating antiferromagnets with a magnitude that may be possible to measure in THz optical experiments. The direction of spin current is determined by the polarization of the optic beam, which is similar to the photogalvanic effect in metals. In our analysis, we assumed a Zeeman magnon coupling with the electromagnetic field, as used in cavity spintronics \cite{Zhang2014a,Cao2015,Yao2015,Sharma2017,Johansen2018, Okuma2018}. 
We note that magneto-optical coupling of light and spins is frequently used for exciting ultrafast spin dynamics in antiferromagnetic insulators \cite{Satoh2010,Tzschaschel2017,Satoh2017}. However, we found this mechanism to be irrelevant for generating spin photocurrents. \footnote{For example, in the inverse Faraday effect the electric field $\bm{\mathfrak{E}}$ produces the effective magnetic field $\bm{\mathfrak{B}}_{\mathrm{eff}} \sim \bm{\mathfrak{E}} \times \bm{\mathfrak{E}}^{*}$. However, as we demonstrated in our paper, the spin current does not appear in the linear order response to $\bm{\mathfrak{B}}_{\mathrm{eff}}$.} 
We also demonstrated that in the presence of asymmetric Dzyaloshinskii-Moriya interactions it is possible to induce spin photocurrents using linearly polarized light. The geometric contribution to the spin current from the magnonic Aharonov–Casher phase has been demonstrated for an example of a honeycomb lattice with staggered Dzyaloshinskii-Moriya interaction.

\begin{acknowledgments}
The authors thank Takuya Satoh for fruitful discussion.  This work was supported by a Grant-in-Aid for Scientific Research (B) (No. 17H02923) and (S) (No. 25220803) from the MEXT of the Japanese Government, JSPS Bilateral Joint Research Projects (JSPS-FBR), and the JSPS Core-to-Core Program, A. Advanced Research Networks. I.P. acknowledges financial support by Ministry of Education and Science of the Russian Federation, Grant No. MK-1731.2018.2 and by Russian Foundation for Basic Research (RFBR), Grant 18-32-00769(mol\_a). A.S.O. acknowledge funding by the RFBR, Grant 17-52-50013, and  the Foundation for the Advancement to Theoretical Physics and Mathematics BASIS Grant No. 17-11-107, and by the Government of the Russian Federation Program 02.A03.21.0006. RLS acknowledges the support of the Natural Sciences and Engineering Research Council of Canada (NSERC). Cette recherche a été financée par le Conseil de recherches en sciences naturelles et en génie du Canada (CRSNG).

\end{acknowledgments}

\appendix
\onecolumngrid

\section{Derivation of magnon spin current}
\label{ap1}
The Fourier transform of the local magnon density is defined as follows
\begin{equation} \label{eq55}
n(\bm{r}_{i}) = \sum_{\bm{\delta}} b^{\dag}_{i+\delta}b_{i+\delta} - a^{\dag}_{i}a_{i} = 
\frac{1}{N}\sum_{\bm{k}\bm{q}\bm{\delta}} e^{-i\bm{q}\cdot\bm{r_{i}}} \left[e^{-i\bm{q}\cdot\bm{\delta}}
b^{\dag}_{\bm{k} + \bm{q}} b_{\bm{k}} - a^{\dag}_{\bm{k} + \bm{q}} a_{\bm{k}} \right],
\end{equation}
where $ \bm{\delta} $ connects the $i$th site on $ A $ sublattice with neighboring sites on $ B $ sublattice. To obtain the equation of motion for $n(\bm{r}_{i})$, we use the Heisenberg equations of motion for $ a_{\bm{k}} $ and $ b_{\bm{k}} $ 
\begin{align}
\dot{a}_{\bm{k}} &= \frac{i}{\hbar}[\mathcal{H},a_{\bm{k}}] = -\frac{i}{\hbar}\left(A_{\bm{k}} a_{\bm{k}} + B^{*}_{\bm{k}} b^{\dag}_{-k}\right), \\
\dot{b}_{-\bm{k}} &= \frac{i}{\hbar}[\mathcal{H},b_{-\bm{k}}] = -\frac{i}{\hbar}\left(A_{\bm{k}} b_{-\bm{k}} + B^{*}_{\bm{k}} a^{\dag}_{k}\right),
\end{align}
where $\mathcal{H}$ is given by Eq.~\eqref{Hm1}, which  gives us (after we change $\bm{k} \to -\bm{k}$ and $-\bm{k}+\bm{q} \to -\bm{k}$ in the first term of Eq.~\eqref{eq55})
\begin{multline}
\frac{\partial n(\bm{r}_{i})}{\partial t} = -\frac{i}{\hbar N} \sum_{\bm{k}\bm{q}\bm{\delta}} e^{-i\bm{q}\cdot\bm{r_{i}}}
\left\lbrace 
\left(A_{\bm{k} + \bm{q}} - A_{\bm{k}}\right)\left[e^{-i\bm{q}\cdot\bm{\delta}}b^{\dag}_{-\bm{k}}b_{-\bm{k}-\bm{q}} + a^{\dag}_{\bm{k} + \bm{q}}a_{\bm{k}}\right] 
\right.
\\
\left.
+ \left(B_{\bm{k} + \bm{q}} - B_{\bm{k}}e^{-i\bm{q}\cdot\bm{\delta}}\right)a_{\bm{k}}b_{-\bm{k}-\bm{q}}
+ \left(e^{-i\bm{q}\cdot\bm{\delta}}B^{*}_{\bm{k} + \bm{q}} - B^{*}_{\bm{k}}\right)a^{\dag}_{\bm{k}+\bm{q}}b^{\dag}_{-\bm{k}}
\right\rbrace.
\end{multline}
In the case when $ \sum_{\bm{\delta}}\bm{\delta} = 0 $,  the lattice form factors do not contribute in the long-wave-length limit  $\bm{q}\to 0$, and we obtain $\partial n_{\bm{q}}/\partial t + i \bm{q} \cdot \bm{J}_{s} = 0$, where $ \bm{J}_{s} $ is given by Eq.~\eqref{JJs}.

\section{Second-order response}
\label{ap2}
In order to calculate the spin current generated by the electromagnetic wave, we use the response theory (see e.~g. [\onlinecite{Tiablikov2013}]). It is easy to show that both the equilibrium magnon spin current and the first order response  vanish.  The second-order response term to $ \mathcal{H}_{I} $ is expressed as 
\begin{equation} \label{eq67}
\langle\bm{J}_{s}(t)\rangle = 
-\sum_{\omega_{1}\omega_{2}} \int_{-\infty}^{t} dt_{1} \int_{-\infty}^{t_{1}} dt_{2}
e^{\epsilon(t_{1} + t_{2} - t)} e^{i\omega_{1}t_{1} + i\omega_{2}t_{2}}
\left\langle
\left[\left[\tilde{\bm{J}}_{s}(t),\tilde{\mathcal{H}}_{I}^{(\omega_{1})}(t_{1})\right],\tilde{\mathcal{H}}_{I}^{(\omega_{2})}(t_{2})\right]
\right\rangle,
\end{equation}
where the operators are in the Heisenberg picture
$\tilde{\bm{J}}_{s}(t) = e^{i\mathcal{H}t} \bm{J}_{s} e^{-i\mathcal{H}t}$,  $\tilde{\mathcal{H}}_{I}^{(\omega)}(t) = e^{i\mathcal{H}t} \mathcal{H}_{I}^{(\omega)} e^{-i\mathcal{H}t}$, and $\mathcal{H}_{I}^{(\omega)}$ is defined by
\begin{equation}
\mathcal{H}_{I}^{(\omega)} = -\frac{1}{2}\sum_{\bm{k}}\left[ h^{(-)}_{\bm{k}}(\omega)\left(M_{\bm{k}}^{*} \alpha_{\bm{k}} + M_{\bm{k}} \beta^{\dag}_{-\bm{k}}\right) + h^{(+)}_{-\bm{k}}(\omega)\left(M_{\bm{k}}\alpha_{\bm{k}}^{\dag} + M_{\bm{k}}^{*}\beta_{-\bm{k}}\right)\right],
\end{equation}
where $ h_{\bm{k}}^{(\pm)}(t) = \sum_{\omega} e^{i\omega t} h_{\bm{k}}^{(\pm)}(\omega) $ are the Fourier components of the magnetic field, which satisfy the identity $ \left [h_{\bm{k}}^{(-)}(-\omega)\right ]^{*} = h_{-\bm{k}}^{(+)}(\omega) $.

Using the Heisenberg equations of motion  for $\alpha_{\bm{k}}$ and $\beta_{-\bm{k}}$, we find
\begin{equation}
\tilde{\mathcal{H}}_{I}^{(\omega)} = -\frac{1}{2}\sum_{\bm{k}}\left[ h^{(-)}_{\bm{k}}(\omega)\left(M_{\bm{k}}^{*}e^{-i\varepsilon_{\bm{k}}t} \alpha_{\bm{k}} + M_{\bm{k}} e^{i\varepsilon_{\bm{k}}t}\beta^{\dag}_{-\bm{k}}\right) + h^{(+)}_{-\bm{k}}(\omega)\left(M_{\bm{k}}e^{i\varepsilon_{\bm{k}}t}\alpha_{\bm{k}}^{\dag} + M_{\bm{k}}^{*}e^{-i\varepsilon_{\bm{k}}t}\beta_{-\bm{k}}\right)\right],
\end{equation}
and
\begin{equation}
\tilde{\bm{J}}_{s} = \sum_{\bm{k}}\left[ 
\bm{\nabla}_{\bm{k}} \varepsilon_{\bm{k}}\left( 
\alpha^{\dag}_{\bm{k}} \alpha_{\bm{k}} + 
\beta^{\dag}_{-\bm{k}} \beta_{-\bm{k}}
\right)
+ \bm{K}_{\bm{k}} e^{-2i\varepsilon_{\bm{k}}t} \alpha_{\bm{k}} \beta_{-\bm{k}}
+ \bm{K}_{\bm{k}}^{*} e^{2i\varepsilon_{\bm{k}}t} \alpha^{\dag}_{\bm{k}} \beta^{\dag}_{-\bm{k}}
\right].
\end{equation}
The commutators in Eq.~\eqref{eq67} can be calculated straightforwardly 
\begin{multline} \label{c1}
\left[\tilde{\bm{J}}(t),\tilde{\mathcal{H}}_{I}^{(\omega_{1})}(t_{1})\right] = -\frac{1}{2}\sum_{\bm{k}} \left\lbrace
-\bm{\nabla}_{\bm{k}} \varepsilon_{\bm{k}}h^{(-)}_{\bm{k}}(\omega_{1})\left[
M_{\bm{k}}^{*}e^{-i\varepsilon_{\bm{k}}t_{1}} \alpha_{\bm{k}} - M_{\bm{k}} e^{i\varepsilon_{\bm{k}}t_{1}}\beta^{\dag}_{-\bm{k}} \right] \right.
\\
\left .
+ \bm{\nabla}_{\bm{k}} \varepsilon_{\bm{k}}h^{(+)}_{-\bm{k}}(\omega_{1})\left[
M_{\bm{k}}e^{i\varepsilon_{\bm{k}}t_{1}} \alpha_{\bm{k}} - M_{\bm{k}}^{*} e^{-i\varepsilon_{\bm{k}}t_{1}}\beta_{-\bm{k}} \right] 
+ \bm{K}_{\bm{k}} e^{-2i\varepsilon_{\bm{k}}t} 
M_{\bm{k}} e^{i\varepsilon_{\bm{k}}t_{1}}
\left[h^{(-)}_{\bm{k}}(\omega_{1}) \alpha_{\bm{k}} + h^{(+)}_{-\bm{k}}(\omega_{1}) \beta_{-\bm{k}}\right] \right .
\\
\left.
- \bm{K}^{*}_{\bm{k}} e^{2i\varepsilon_{\bm{k}}t} 
M^{*}_{\bm{k}} e^{-i\varepsilon_{\bm{k}}t_{1}}
\left[h^{(-)}_{\bm{k}}(\omega_{1}) \beta^{\dag}_{-\bm{k}} + h^{(+)}_{-\bm{k}}(\omega_{1}) \alpha^{\dag}_{\bm{k}}\right]
\right\rbrace,
\end{multline}
and
\begin{multline}
\left[\left[\tilde{\bm{J}}(t),\tilde{\mathcal{H}}_{I}^{(\omega_{1})}(t_{1})\right],\tilde{\mathcal{H}}_{I}^{(\omega_{2})}(t_{2})\right] 
\\= \frac{1}{4}\sum_{\bm{k}}
\left\lbrace\left [
-\bm{\nabla}_{\bm{k}} \varepsilon_{\bm{k}}|M_{k}|^{2}\left(
e^{-i\varepsilon_{k}(t_{1} - t_{2})} + e^{i\varepsilon_{k}(t_{1} - t_{2})}
\right)+
\bm{K}_{\bm{k}}M_{\bm{k}}^{2}e^{i\varepsilon_{\bm{k}}(t_{1}+t_{2} - 2t)}
+ \bm{K}^{*}_{\bm{k}}M_{\bm{k}}^{*2}e^{-i\varepsilon_{\bm{k}}(t_{1}+t_{2} - 2t)}
\right ]
\right .
\\
\left .
\times\left (
h^{(-)}_{ \bm{k}}(\omega_{1}) h^{(+)}_{-\bm{k}}(\omega_{2})+
h^{(+)}_{-\bm{k}}(\omega_{1}) h^{(-)}_{ \bm{k}}(\omega_{2})
\right )
\right\rbrace.
\end{multline}
The integration over $ t_{1} $ and $ t_{2} $ in Eq.~\eqref{eq67} is performed as follows
\begin{align}
\int_{-\infty}^{t} dt_{1} \int_{-\infty}^{t_{1}} dt_{2} 
e^{\epsilon(t_{1} + t_{2} -t)} e^{i\omega_{1}t_{1} + i\omega_{2}t_{2}} e^{-i\varepsilon_{\bm{k}}(t_{1} - t_{2})} & = 
-\frac{e^{i(\omega_{1} + \omega_{2})t + \epsilon t}}{(\varepsilon_{k} + \omega_{2} - i\epsilon)(\omega_{1} + \omega_{2} -2i\epsilon)}, \\
\int_{-\infty}^{t} dt_{1} \int_{-\infty}^{t_{1}} dt_{2} 
e^{\epsilon(t_{1} + t_{2} -t)} e^{i\omega_{1}t_{1} + i\omega_{2}t_{2}} e^{i\varepsilon_{\bm{k}}(t_{1} - t_{2})} & = 
\frac{e^{i(\omega_{1} + \omega_{2})t + \epsilon t}}{(\varepsilon_{k} - \omega_{2} + i\epsilon)(\omega_{1} + \omega_{2} -2i\epsilon)}, \\
\int_{-\infty}^{t} dt_{1} \int_{-\infty}^{t_{1}} dt_{2} 
e^{\epsilon(t_{1} + t_{2} -t)} e^{i\omega_{1}t_{1} + i\omega_{2}t_{2}} e^{i\varepsilon_{\bm{k}}(t_{1} + t_{2} - 2t)} & = 
\frac{-e^{i(\omega_{1} + \omega_{2})t + \epsilon t}}{(\varepsilon_{k} + \omega_{2} - i\epsilon)(\omega_{1} + \omega_{2} + 2\varepsilon_{k} - 2i\epsilon)}, \\
\label{t1}
\int_{-\infty}^{t} dt_{1} \int_{-\infty}^{t_{1}} dt_{2} 
e^{\epsilon(t_{1} + t_{2} -t)} e^{i\omega_{1}t_{1} + i\omega_{2}t_{2}} e^{-i\varepsilon_{\bm{k}}(t_{1} + t_{2} - 2t)} & = 
\frac{-e^{i(\omega_{1} + \omega_{2})t + \epsilon t}}{(\varepsilon_{k} - \omega_{2} + i\epsilon)(2\varepsilon_{k} - \omega_{1} - \omega_{2} +  2i\epsilon)}.
\end{align}

Combining togeter Eqs~(\ref{c1})--(\ref{t1}), we obtain the following expression for the magnon spin current
\begin{multline}
\langle\bm{J}_{s}(t)\rangle = 
\frac{1}{4}\sum_{\omega_{1}\omega_{2} \bm{k}} e^{i(\omega_{1} + \omega_{2})t + \epsilon t}
\left\lbrace
\frac{|M_{\bm{k}}|^{2}\bm{\nabla}_{\bm{k}}\varepsilon_{k}}{\omega_{1}+\omega_{2}-2i\epsilon}\left(\frac{1}{\varepsilon_{\bm{k}} - \omega_{2} + i\epsilon} -  \frac{1}{\varepsilon_{\bm{k}} + \omega_{2} - i\epsilon}\right)
\right .
\\
\left .
+\frac{\bm{K}_{\bm{k}} M^{2}_{\bm{k}}}{(\varepsilon_{\bm{k}} +\omega_{2}-i\epsilon)(2\varepsilon_{\bm{k}} + \omega_{1} + \omega_{2} -2i\epsilon)} 
+\frac{\bm{K}^{*}_{\bm{k}} M^{*2}_{\bm{k}}}{(\varepsilon_{\bm{k}} -\omega_{2}+i\epsilon)(2\varepsilon_{\bm{k}} - \omega_{1} - \omega_{2} +2i\epsilon)}
\right\rbrace \\
\times \left [
h^{(-)}_{ \bm{k}}(\omega_{1}) h^{(+)}_{-\bm{k}}(\omega_{2})+
h^{(+)}_{-\bm{k}}(\omega_{1}) h^{(-)}_{ \bm{k}}(\omega_{2})
\right ].
\end{multline}
By changing, $\omega_{2} \to -\omega_{2}$ ($\omega_{1} \to -\omega_{1}$) in the term proportional to $h^{(-)}_{ \bm{k}}(\omega_{1}) h^{(+)}_{-\bm{k}}(\omega_{2})$ ($h^{(+)}_{-\bm{k}}(\omega_{1}) h^{(-)}_{ \bm{k}}(\omega_{2})$),  we can rewrite the expression above in the manifestly real form
\begin{multline}
\langle\bm{J}_{s}(t)\rangle = 
\frac{1}{4}\sum_{\omega_{1}\omega_{2} \bm{k}} e^{i(\omega_{1} - \omega_{2})t + \epsilon t}
\left\lbrace
\frac{|M_{\bm{k}}|^{2}\bm{\nabla}_{\bm{k}}\varepsilon_{k}}{\omega_{1}-\omega_{2}-2i\epsilon}\left(\frac{1}{\varepsilon_{\bm{k}} + \omega_{2} + i\epsilon} -  \frac{1}{\varepsilon_{\bm{k}} - \omega_{2} - i\epsilon}\right)
\right .
\\
\left .
+\frac{\bm{K}_{\bm{k}} M^{2}_{\bm{k}}}{(\varepsilon_{\bm{k}} -\omega_{2}-i\epsilon)(2\varepsilon_{\bm{k}} + \omega_{1} - \omega_{2} -2i\epsilon)} 
+\frac{\bm{K}^{*}_{\bm{k}} M^{*2}_{\bm{k}}}{(\varepsilon_{\bm{k}} +\omega_{2}+i\epsilon)(2\varepsilon_{\bm{k}} - \omega_{1} + \omega_{2} +2i\epsilon)}
\right\rbrace \\
\times 
h^{(-)}_{ \bm{k}}(\omega_{1}) h^{(-)*}_{\bm{k}}(\omega_{2})
\\
+\frac{1}{4}\sum_{\omega_{1}\omega_{2} \bm{k}} e^{-i(\omega_{1} - \omega_{2})t + \epsilon t}
\left\lbrace
\frac{|M_{\bm{k}}|^{2}\bm{\nabla}_{\bm{k}}\varepsilon_{k}}{\omega_{1}-\omega_{2}+2i\epsilon}\left(\frac{1}{\varepsilon_{\bm{k}} + \omega_{2} - i\epsilon} -  \frac{1}{\varepsilon_{\bm{k}} - \omega_{2} + i\epsilon}\right)
\right .
\\
\left .
+\frac{\bm{K}_{\bm{k}} M^{2}_{\bm{k}}}{(\varepsilon_{\bm{k}} +\omega_{2}-i\epsilon)(2\varepsilon_{\bm{k}} - \omega_{1} + \omega_{2} -2i\epsilon)} 
+\frac{\bm{K}^{*}_{\bm{k}} M^{*2}_{\bm{k}}}{(\varepsilon_{\bm{k}} -\omega_{2}+i\epsilon)(2\varepsilon_{\bm{k}} + \omega_{1} - \omega_{2} +2i\epsilon)}
\right\rbrace \\
\times 
h^{(-)*}_{ \bm{k}}(\omega_{1}) h^{(-)}_{\bm{k}}(\omega_{2}).
\end{multline}

If we take the diagonal part of this expression at $\omega_{1} = \omega_{2} \to \omega_{\bm{k}}$, we obtain time-independent component of the spin current in Eq.~\eqref{JJss3}.

\twocolumngrid

\bibliography{afmspin}
\end{document}